\documentstyle[aasms4,12pt]{article} 

\begin{document}
\title{ANALYTIC MODEL FOR ADVECTION-DOMINATED ACCRETION FLOWS \\
       IN A GLOBAL MAGNETIC FIELD}
\author{\sc Osamu Kaburaki}
\vspace {1cm}
\affil{Astronomical Institute, Graduate School of Science, 
Tohoku University, \\
Aoba-ku, Sendai, 980-8578, Japan; okabu@astr.tohoku.ac.jp}

\begin{abstract}
A model for advection-dominated accretion flows (ADAFs) in a 
global magnetic field is proposed. In contrast to the well 
known ADAF models in which the viscosity of a fluid determines both angular 
momentum transfer and energy dissipation in the flow, the magnetic field 
and the electric resistivity, respectively, control them in this model. 
A manageable set of analytic solutions for the flow and the magnetic field 
is obtained to vertically non-integrated basic equations. This set 
describes mathematically a fully advective accretion flow and, in physically 
plausible situations for most AGNs, it is also confirmed that the radiation 
cooling estimated on this solution is really negligible compared with 
the internal energy of the flow. 

\end{abstract}

\keywords{accretion, accretion disks --- magnetohydrodynamics: 
MHD --- galaxies: active --- black hole physics}

\section{INTRODUCTION} 

In recent years, advection-dominated accretion flows (ADAFs) are drawing 
much attention in astrophysics because the radiation spectrum of Sgr A$^*$ is 
well reproduced in the framework of such models (Narayan, Yi \& Mahadevan 
1995; Manmoto, Mineshige \& Kusunose 1997; Narayan et al. 1998). The attempts 
are further made to apply them also to some other AGNs and soft X-ray 
transients (for a review, see Narayan, Mahadevan \& Quataert 1999).  Such 
a flow may be realized when the rate of mass accretion is smaller than 
the Eddington accretion rate, and the flow becomes optically thin while 
geometrically thick. Almost whole energy dissipated in the flow is advected 
down to the central object owing to an inefficient radiation cooling. 

The basic ideas of such ADAF models has been developed by a number of 
researchers (e.g., Ichimaru 1977; Rees et al. 1982; Narayan \& Yi 1994, 
1995; Abramowicz et al. 1995). We call them the ``viscous" ADAF models 
because both angular-momentum transfer and energy dissipation in the flow 
is undertaken by the kinematic viscosity whose size is usually specified 
by the $\alpha$ parameter: $\nu = (2/3)\alpha C_{\rm S}H$, where 
$C_{\rm S}$ is the sound velocity and $H$ is the half-thickness of 
the flow. 
Another parameter $\beta$ enters into this model through the assumed 
presence of turbulent magnetic fields (Narayan \& Yi 1995): 
$p_{\rm m} = 3(1-\beta)\ C_{\rm S}^{\ 2} \rho$, where $p_{\rm m}$ is the 
turbulent magnetic pressure and $\rho$ is the mass density. 
The spectrum of Sgr A$^*$ in a range from radio up to 
hard X-ray has been explained by the synchrotron emission from high 
temperature electrons, its inverse-Compton scattered component and 
bremsstrahlung. For the presence of synchrotron emission, the inclusion 
of turbulent magnetic field is crucial in this model. 

Although the above model completely ignores the presence of an ordered 
magnetic field, it is natural to expect the presence of such fields in 
most AGNs and especially in the central region of our Galaxy (e.g., 
Yusef-Zadeh, Morris \& Chance 1984). Taking this fact into account, we 
propose another ADAF model which shall be called the ``resistive" ADAF 
model in order to distinguish it from the viscous ADAF. This name reflects 
the facts that angular momentum is transported by a global magnetic field 
and energy is released as a resistive dissipation of the electric current 
driven by a rotational motion of the accreting plasma. Originally, this 
model appeared (Kaburaki 1986, 1987) as a magnetic counterpart of the 
standard model (usually called the $\alpha$-disk model) for optically 
thick, geometrically thin accretion disks (Shakura \& Sunyaev 1973). 

Analogously to the $\alpha$-disk model, this model contains a parameter 
$\Delta$ which specifies the efficiency of dissipation: $\sigma^{-1} \propto 
\Delta^2$, where $\sigma^{-1}$ is the electric resistivity. The parameter 
$\Delta$ in this model also has a geometrical meaning of the half 
opening-angle of a disk, which is assumed for simplicity to be constant 
throughout the disk. The set of analytic solutions for the flow and the 
magnetic field was obtained there under the assumption of geometrically 
thin disk ($\Delta \ll 1$), and it describes an accretion flow which has 
a reduced Keplerian rotation and a magnetically confined vertical structure. 

Although the heating process in the disk has been discussed in the above 
papers, the cooling process has not been discussed and implicitly assumed 
to be in local balance with the heating. Since the energy equation 
is not included explicitly, the solution may be inconsistent from a 
viewpoint of energy transport. In principle, the fraction of advected part 
of the released energy can be calculated from that set of solutions. 
Actually, however, the inclusion of a vertical motion in the model of 
Kaburaki (1987, hereafter referred to as K\ 87) has made this task rather 
difficult and obscure. Therefore, in the present paper, we intend to obtain 
a more manageable set of solutions to a similar problem by adding further 
simplifying assumptions of no vertical flows and electric currents. Although 
such a solution has already been obtained in a previous paper (Kaburaki 1986), 
it is unsatisfactory because the effect of pressure gradient has been 
omitted without justification. 

Our new solution obtained in the present paper turns out {\it a posteriori} 
to describe a fully advective accretion flow (as a preliminary report, 
see Kaburaki 1999). In other words, the requirements for dynamical balances 
and the geometry of a flow specify in our model also the way of energy 
transport. The consistency of this solution as an ADAF model in most 
physical situations expected in typical AGNs is also confirmed. Namely, 
it is shown by using this solution that the flow is actually optically 
thin and the radiation cooling is negligible compared with the heating. 
Therefore, our resistive ADAF model may be useful in calculating spectra 
form various AGNs and in considering stabilities of realistic ADAFs. 
Some results in such applications will appear elsewhere.

This paper is organized as follows. In \S 2, a set of resistive MHD 
equations is first introduced as our starting point, and then various 
assumptions to simplify these equations are stated. The resulting equations 
in spherical polar coordinates are cited in \S 3, and the effects of the 
individual assumptions and the policy for solving these equations are 
discussed. A set of approximate solutions which we propose as a model of 
resistive ADAFs is written out in \S 4 and its properties, including energy 
budget, are examined. \S 5 is devoted to determining the disk edges which 
are defined conceptually as the limiting radii for the validity of 
this solution. In \S 6, various quantities predicted by this model 
are suitably scaled to typical situations in most AGNs and the physical 
consistency of this solution as an ADAF model is also checked. 
Finally, main issues which may be raised about our model are discussed 
in \S 7. 
Some remarks concerning a relation between the rotation law and 
geometry of a disk are also described in Appendix.

\section{BASIC EQUATIONS AND SIMPLIFYING ASSUMPTIONS}

The basic equations adopted in the resistive ADAF model are those of resistive 
MHD which are written in usual notation as follows: 
  \begin{equation} 
        \frac{\partial \rho}{\partial t} 
           + {\bf \nabla}\cdot(\rho{\bf v}) = 0, 
  \end{equation}
  \begin{equation}
        \frac{\partial {\bf v}}{\partial t} 
           + ({\bf v}\cdot{\bf \nabla}){\bf v} = 
           - \frac{1}{\rho}\ {\bf \nabla}p 
           - \frac{1}{4\pi\rho}\ [{\bf B}\times({\bf \nabla}\times{\bf B})] 
           + {\bf g}, \qquad (\nu = 0) 
  \label{eqn:eqmom}
  \end{equation}
  \begin{equation} 
        {\bf \nabla}\cdot{\bf B} = 0, 
  \end{equation}
  \begin{equation} 
        \frac{\partial {\bf B}}{\partial t} = 
            {\bf \nabla}\times({\bf v}\times{\bf B}) 
            + \frac{c^2}{4\pi\sigma}\ \triangle{\bf B}. 
            \qquad (\sigma = \mbox{const.}) 
  \label{eqn:eqind}
  \end{equation}
In the equation of motion (\ref{eqn:eqmom}), the viscosity of the fluid 
is completely neglected in order to make the contrast between the resistive 
and viscous ADAF models clear. Instead, the term of magnetic diffusion is 
retained in the induction equation (\ref{eqn:eqind}) assuming the constancy 
of electrical conductivity $\sigma$ for simplicity. The above set of 
equations forms an apparently closed set in the sense that the number of 
equations and unknowns are the same. In fact, however, the number of 
equations is insufficient by one. This is because Maxwell's equations 
guarantee that ${\rm div}\ {\bf B}=0$ always holds once it is satisfied 
initially. This point will become more clear in \S 3. Other related 
quantities of our interest are calculated from the following subsidiary 
equations: the current density, electric field and charge density, from 
  \begin{equation}
       {\bf j} = \frac{1}{4\pi}\ {\bf \nabla}\times{\bf B}, 
            \quad {\bf E} = \frac{\bf j}{\sigma} 
            - \frac{1}{c}\ {\bf v}\ \times{\bf B},
            \quad q = \frac{1}{4\pi}\ {\bf \nabla}\cdot{\bf E},
  \end{equation}
respectively. The temperature of the fluid is simply assumed to be common 
to electrons and ions, and is calculated from the ideal gas law 
  \begin{equation}
        p = \frac{R}{\bar{\mu}}\ \rho T,
  \end{equation}
neglecting the radiation pressure ($\bar{\mu}$ is the mean molecular 
weight and $R$ is the gas constant). 

Hereafter, we employ spherical polar coordinates ($r$, $\theta$, $\varphi$) 
since the gravity is spherically symmetric. The main assumptions used in 
simplifying the above basic equations are, 1) stationarity, 2) axisymmetry, 
3) geometrically thin disk, 4) well-developed magnetic disk, 5) reduced 
Keplerian rotation and 6) vanishing of $v_{\theta}$ and $j_{\theta}$. All 
these assumptions except the last one are the same as employed in K\ 87. 
Although the solution has been obtained there without 6), we add it here 
in order to get a more manageable set of solutions. 

The first two of the above assumptions imply that $\partial/\partial t = 0$ 
and $\partial/\partial{\varphi} = 0$ in every component equation. The most 
essential assumption is the third one which demands that $\Delta \ll 1$. 
This fact allows us to treat $\Delta$ as a smallness parameter in the 
following discussion. In such situations, a derivative with respect to 
$\theta$ results in a large quantity (introduce $\xi\equiv 
(\theta-\pi/2)/\Delta$ then $\partial/\partial\theta = 
\Delta^{-1}\partial/\partial\xi$). It is worth emphasizing, however, that 
the thin disk condition is satisfied even by rather geometrically thick 
disks since 1 radian is about 60$^{\circ}$ of arc. 

The fourth assumption states that, when the total magnetic field is divided 
into the externally-given seed field ${\bf B}_0$ and the component produced 
in the disk {\bf b}, we expect $\vert{\bf b}\vert \gg \vert{\bf B_0}\vert$ 
inside the disk except near a disk edge where $\vert{\bf b}\vert \sim 
\vert{\bf B_0}\vert$. The actual implication of the fifth assumption is 
that of large magnetic Reynolds numbers $\Re$ (strictly speaking, 
$\Re^2(r) \gg 1$) in the disk except near its inner edge. 
Namely, as it will tern out {\it a posteriori} from the resulting 
solutions, the assumption of reduced Keplerian rotation 
($v_{\varphi}=\mbox{const.}\times v_{\rm K}\ <\ v_{\rm K}$, 
where $v_{\rm K}\equiv\sqrt{GM/r}$ represents the Kepler velocity) is 
justified when $\Re^2(r)\gg 1$ (see \S 4). The sixth assumption results 
in an $r$-independent accretion rate $\dot{M}$ and the dependence 
$b_{\phi}\propto r^{-1}$.

\section{LEADING ORDER EQUATIONS IN $\Delta$}

The set of basic equations is simplified first according to the above 
assumptions except 5) and 6). Since the flow is assumed to be geometrically 
thin, only the leading order terms in $\Delta$ are retained in each component 
of the basic equations. Regarding $b_{r}$, $b_{\varphi}$, $v_{r}$, 
$v_{\varphi}$, $\rho$ and $p$ as quantities of order unity in $\Delta$, 
we obtain the following component equations, within the approximation 
$\sin \theta \sim 1$. \\{\bf mass continuity}:
\begin{equation}
  \frac{\partial}{\partial r}(r^2\rho v_r) 
  + \frac{r}{\Delta}\ \frac{\partial}{\partial \xi}(\rho v_{\theta}) = 0.
\end{equation}
{\bf magnetic flux conservation}:
\begin{equation}
  \frac{\partial}{\partial r}(r^2 b_r) 
   + \frac{r}{\Delta}\ \frac{\partial b_{\theta}}{\partial \xi} = 0.
\end{equation}
{\bf equation of motion}:
  \begin{eqnarray}
      \lefteqn{\left[ v_r\ \frac{\partial}{\partial r} 
         + \frac{v_{\theta}}{\Delta r}\ \frac{\partial}{\partial \xi} 
         \right]v_r - \frac{v_{\varphi}^2}{r}} \nonumber \\
       & & = -\frac{1}{\rho}\ \frac{\partial p}{\partial r} 
       -\frac{GM}{r^2} + \frac{1}{4\pi\rho r} \left[ 
       \frac{b_{\theta}}{\Delta}\ \frac{\partial b_r}{\partial \xi} 
       -b_{\varphi}\ \frac{\partial}{\partial r}(rb_{\varphi}) \right],
  \label{eqn:momr}
  \end{eqnarray}
\begin{equation}
  \frac{\partial p}{\partial \xi} 
   + \frac{1}{8\pi}\ \frac{\partial}{\partial\xi}(b_r^{\ 2}+b_{\varphi}^{\ 2})
  = 0,
\end{equation}
  \begin{equation}
        \left[v_r\ \frac{\partial}{\partial r}
        +\frac{v_{\theta}}{\Delta r}\ \frac{\partial}{\partial\xi}\right]
        v_{\varphi} + \frac{v_{\varphi}v_r}{r} 
  = \frac{1}{4\pi\rho r}\left[ b_r\ \frac{\partial}{\partial r}(rb_{\varphi})
        +\frac{b_{\theta}}{\Delta}\ \frac{\partial b_{\varphi}}{\partial\xi}
        \right]. 
  \end{equation}
{\bf induction equation}:
\begin{equation}
  \frac{\partial}{\partial\xi}\left[ v_r b_{\theta} - v_{\theta}b_r 
   + \frac{c^2}{4\pi\sigma\Delta} \frac{1}{r} 
     \ \frac{\partial b_r}{\partial\xi} \right] = 0,
\end{equation}
\begin{equation}
  \frac{\partial}{\partial r} \left[ r\left( v_r b_{\theta} - v_{\theta}b_r 
   + \frac{c^2}{4\pi\sigma\Delta} \frac{1}{r} 
     \ \frac{\partial b_r}{\partial\xi} \right)\right] = 0,
\end{equation}
\begin{equation}
  \frac{\partial}{\partial r}\left[ r(v_{\varphi}b_r 
       - v_r b_{\varphi}) \right] 
  +\frac{1}{\Delta}\ \frac{\partial}{\partial \xi} 
   \left[ v_{\varphi}b_{\theta} - v_{\theta}b_{\varphi} 
   + \frac{c^2}{4\pi\sigma\Delta} \frac{1}{r}\ \frac{\partial b_{\varphi}}
       {\partial\xi} \right] = 0.
  \end{equation}
From the equations of mass continuity and flux conservation, it turns out 
that $v_{\theta}$ and $b_{\theta}$ are quantities of order $\Delta$. 
Further, the assumption of $v_{\theta}=0$ results in an $r$-independent 
radial mass flux, $r^2\rho v_r$. 

At this stage of approximation, there remain so many terms in the 
$r$-component of the equation of motion. Only the last term on the 
the right-hand side can be dropped by a further assumption of $j_{\theta}=0$ 
(for the component expressions of the current density to the leading order 
in $\Delta$, see K\ 87). As confirmed retrospectively, however, $\rho,
\ p\propto\Re^2$ while $v_r \propto \Re^{-1}$. Therefore, when $\Re^2\gg 1$ 
(the assumption 5), equation (\ref{eqn:momr}) reduces to 
  \begin{equation}
         \frac{GM}{r^2}=\frac{v_{\varphi}^2}{r}
         -\frac{1}{\rho}\ \frac{\partial p}{\partial r}. 
  \label{eqn:Kepl}
  \end{equation} 
Owing to the presence of the outward pressure-gradient, the effect of 
gravity is somewhat reduced. Anticipating the same $r$- and 
$\theta$-dependences for the gravity and pressure terms, we have a reduced 
Keplerian rotation from this equation. The validity of this anticipation 
is again justified {\it a posteriori} from the resulting solution. 

The $\theta$-component of the equation of motion can be integrated to 
give the ``vertical" pressure balance for the disk: 
  \begin{equation} 
        p + \frac{b_{\varphi}^2}{8\pi}=\tilde{p}(r),
  \end{equation}
where $\tilde{p}(r)$ represent a function of $r$ only. In the above 
procedure, the magnetic pressure due to $b_r$ has been omitted in 
expectation of the relation $b_{\varphi}/b_r \propto\Re$, which is also 
confirmed retrospectively. This equation implies the confinement of the 
accreting flow into a disk structure by the induced toroidal magnetic 
field which changes sign at the equatorial plane. The $\varphi$-component 
of the equation of motion, on the other hand, reduces to 
  \begin{equation}
       v_r\frac{\partial(rv_{\varphi})}{\partial r}\
        =\ \frac{1}{4\pi\Delta}\ \frac{b_{\theta}}{\rho}\ 
        \frac{\partial b_{\varphi}}{\partial\xi}
  \end{equation}
when simplified by the assumption 6). This equation describes the transfer 
of angular momentum by the magnetic stress. The extracted angular 
momentum is carried away from the disk along the poloidal lines of force. 

Both $r$- and $\theta$-components of the induction equation reduce to 
  \begin{equation}
     v_r = -\frac{c^2}{4\pi\sigma\Delta}\ \frac{1}{r b_{\theta}}\ 
        \frac{\partial b_r}{\partial\xi},
  \end{equation}
since the quantities in the brackets of these equations are proportional to 
the $\varphi$-component of the electric field $E_{\varphi}$ which should 
vanish in axisymmetric situations (see Ohm's law in K\ 87). Thus, 
the degeneracy of the two equations have become evident. However, the 
degeneracy itself is not a consequence of symmetry but of the nature 
inherent in Maxwell's equations as stated before. We can expect from this 
equation that $v_r\propto\Re^{-1}$. 

As for the $\varphi$-component of the induction equation, it is also 
simplified by the assumption of $v_{\theta}=0$ and by the fact that 
$v_{\varphi}$ is independent of $\xi$ since it is proportional to the 
Kepler velocity. Further eliminating $b_{\theta}$ with the aid of magnetic 
flux conservation, we obtain 
\begin{equation}
  r^2 b_r\ \frac{\partial}{\partial r} \left( \frac{v_{\varphi}}{r} \right)
  - \frac{\partial}{\partial r}(rv_r b_{\varphi}) + \frac{c^2}{4\pi\sigma
    \Delta^2}\frac{1}{r}\ \frac{\partial^2 b_{\varphi}}{\partial\xi^2} = 0.
\end{equation}
In solving this equation, we assume intuitively the proportionality of 
the first and second terms. Afterwards, from the resulting solution, the 
consistency of such an assumption is confirmed. This is the same technique 
as used in solving equation (\ref{eqn:Kepl}). 

\section{AN ADAF SOLUTION}

The above set of partial differential equations is incomplete in the sense 
that the number of equations is insufficient by one. Usually, this shortage 
is supplemented by the energy equation or the polytropic one, and it makes 
the process of solving the set of equations rather difficult. Fortunately, 
however, we can get a solution here without introducing another such 
equation. The set is solved by the method of approximate variable separation. 
Starting from the assumption of $b_{\varphi}\propto\tanh\xi$, we can obtain 
almost automatically the angular dependences of other quantities according 
to the equations. The only approximation needed is to accept the relation 
$({\rm d}^3/{\rm d}\xi^3)\tanh\xi \simeq -2\ {\rm sech}^4\xi$ which hold 
fairly good around the disk's midplane. Including the method of obtaining 
the radial parts of the unknowns, the process is essentially the same 
as described in K\ 87. Therefore, we cite below only the final results. 

The boundary value of the external magnetic field $B_0$ is taken into 
account at a reference radius $r_0$, which may be the inner or outer edge of 
the accretion disk (they will be defined explicitly in the next section). 
Then the set of solutions for various quantities are written as 
\begin{equation}
  b_r(r,\ \xi) =\ \tilde{b}_r(r)\ \mbox{sech}^2\xi\ \tanh\xi, \qquad 
  \tilde{b}_r(r) = B_0\left( \frac{r}{r_0}\right)^{-3/2},
\end{equation}
\begin{equation}
  b_{\theta}(r,\ \xi) =\ \tilde{b}_{\theta}(r)\ \mbox{sech}^2\xi, \qquad 
  \tilde{b}_{\theta}(r) = \frac{\Delta}{4}\ B_0\left( \frac{r}{r_0} 
   \right)^{-3/2},
\end{equation}
  \begin{equation}
     b_{\varphi}(r,\ \xi) =-\tilde{b}_{\varphi}(r)\tanh\xi, \qquad 
   \tilde{b}_{\varphi}(r) = \Re(r_0)\ B_0 
        \left( \frac{r}{r_0} \right)^{-1}, 
  \end{equation}
  \begin{equation}
     v_r(r,\ \xi) =-\tilde{v}_r(r)\ \mbox{sech}^2\xi, \qquad 
   \tilde{v}_r(r) = \frac{v_{\rm K}(r_0)}{\sqrt{3}\ \Re(r_0)} 
        \left( \frac{r}{r_0} \right)^{-1}, 
  \end{equation}
  \begin{equation}
     v_{\theta}(r,\ \xi) = 0,
  \end{equation}
  \begin{equation}
     v_{\varphi}(r,\ \xi) = \tilde{v}_{\varphi}(r), \qquad 
   \tilde{v}_{\varphi}(r) = \frac{v_{\rm K}(r_0)}{\sqrt{3}} 
        \left( \frac{r}{r_0} \right)^{-1/2},
  \end{equation}
  \begin{equation}
     p(r,\ \xi) =\ \tilde{p}(r)\ \mbox{sech}^2\xi, \quad
   \tilde{p}(r) = \frac{\Re^2(r_0)B_0^2}{8\pi} 
        \left( \frac{r}{r_0} \right)^{-2},
  \end{equation}
  \begin{equation}
     \rho(r,\ \xi) =\ \tilde{\rho}(r)\ \mbox{sech}^2\xi, \qquad
   \tilde{\rho}(r) = \frac{3\ \Re^2(r_0)B_0^2}{8\pi\ v_{\rm K}^2(r_0)} 
        \left( \frac{r}{r_0} \right)^{-1},
  \end{equation}
  \begin{equation}
     j_r(r,\ \xi) =-\tilde{j}_r(r)\ \mbox{sech}^2\xi, \qquad
   \tilde{j}_r(r) = \frac{c}{4\pi\Delta}\ \frac{\Re(r_0)B_0}{r_0} 
        \left( \frac{r}{r_0} \right)^{-2},
  \end{equation}
  \begin{equation}
     j_{\theta}(r,\ \xi) = 0,
  \end{equation}
  \begin{equation}
     j_{\varphi}(r,\ \xi) =-\tilde{j}_{\varphi}(r)\ \mbox{sech}^4\xi, \qquad
   \tilde{j}_{\varphi}(r) = \frac{c}{4\pi\Delta}\ \frac{B_0}{r_0} 
        \left( \frac{r}{r_0} \right)^{-5/2},
  \end{equation}
  \begin{equation}
   T(r,\ \xi) \equiv \frac{\bar{\mu}}{R}\ \frac{p(r,\ \xi)}{\rho(r,\ \xi)} 
      = \frac{\bar{\mu}}{R}\ \frac{v_{\rm K}^2(r_0)}{3} 
        \left( \frac{r}{r_0}\right)^{-1}.
  \end{equation}
The definition and its $r$-dependence of the magnetic Reynolds number in 
our model are 
  \begin{equation}
   \Re(r) \equiv \frac{\tilde{b}_{\varphi}}{\tilde{b}_r} 
        = \frac{\tilde{v}_{\varphi}}{\tilde{v}_r} 
        = \Re(r_0) \left( \frac{r}{r_0} \right)^{1/2},
        \qquad \Re(r_0) = \frac{\pi\sigma\Delta^2 r_0\ v_{\rm K}(r_0)}
        {\sqrt{3}\ c^2}. 
  \label{eqn:Re}
  \end{equation}
From the latter expression, it turns out that our magnetic Reynolds number 
is reflecting the vertical structure of the disk since $\Delta r$ and 
$\Delta v_{\rm K}$ represent typical sizes of the height and the velocity 
in the $\theta$-direction, respectively. 

Here, we summarize the characteristic features of the above set of solutions. 
The plasma density and pressure are large quantities of order $\Re^2$, 
reflecting the development of a disk-like structure of the flow. In order 
to vertically support this configuration, a large toroidal magnetic field 
$b_{\varphi}$ of order $\Re$ is required to appear, and this fact also 
guarantees sufficient extraction of angular momentum from the accreting 
matter. Controlled by this extraction, the plasma falls inwardly. This 
inflow is understood also as a result of resistive diffusion across the 
poloidal magnetic field lines. Reflecting this fact, $v_r$ is inversely 
proportional to $\Re$. The presence of a factor $\Re$ in $j_r$ represents 
that the rotation dominated accretion flow acts mainly as a poloidal current 
generator, from an electrodynamical point of view. 

Since the poloidal and toroidal components of vector quantities have different 
$r$-dependences, our solution is not such a similarity solution as obtained 
in the framework of viscous ADAF (Narayan \& Yi 1994). Reflecting the disk's 
geometry, the {\it rotation velocity} is a reduced Keplerian (see Appendix 
for further details) at the points on a surface of constant $r$. However, 
the plane of rotation is always parallel to the disk's midplane so that 
the center of rotation generally does not coincide with the center of 
gravitational attraction. It is also interesting to see that a polytrope-like 
relation holds only for the radial functions, i.e., 
$\tilde{p}\propto\tilde{\rho}^{\ \gamma}$ with $\gamma=2$. The assumption 
of $v_{\theta}=0$ assures an $r$-independent mass accretion rate 
\begin{equation}
  \dot{M} = -4\pi\Delta\ \int_{0}^{\infty}r^2\rho v_r\ {\rm d}\xi 
   = \frac{\Delta\Re(r_0)B_0^{\ 2}r_0^{\ 2}}{\sqrt{3}\ v_{\rm K}(r_0)}. 
\label{eqn:Mdot}
\end{equation}
A similar thing is assured by the assumption of $j_{\theta}=0$ for the total 
poloidal current flowing in the disk. In order to fulfill the requirement of 
poloidal current closure, however, non-zero $j_{\theta}$ should be included. 
Our model, therefore, is shifting this problem to the regions both beyond 
the outer edge and within the inner edge of the disk. 

Based on the above cited solution, we can discuss the energy budget in the 
accretion flow. Since the electric current vector is dominated by its 
$r$-component, we have for the local Joule dissipation rate
  \begin{equation}
    q_{\rm J}^{\ +}(r,\ \xi) 
      \equiv \frac{{\bf j}^2}{\sigma} \sim \frac{j_r^2}{\sigma}
      = \frac{\Re(r_0)}{2\sqrt{3}}\ \frac{B_0^2}{8\pi}\ \Omega_{\rm K}(r_0)
        \left( \frac{r}{r_0} \right)^{-4} \mbox{sech}^4 \xi, 
  \end{equation}
where $\Omega\equiv v_{\varphi}/r$ denotes the angular velocity (within 
the approximation of $\sin\theta \sim 1$) and $\sigma$ has been eliminated 
by equation (\ref{eqn:Re}). Although the origin of the resistivity need 
not be specified in our model, it is very likely to be of anomalous type 
since a large current density in the disk may cause a plasma turbulence. 
On the other hand, the advection cooling, which is defined as 
  \begin{equation}
    q_{\rm adv}^{\ -} \equiv \rho T\ ({\bf v}\cdot\mbox{grad})s 
      = \frac{1}{r^2}\ \frac{\partial}{\partial r}\ [r^2\rho hv_r] 
      - v_r\frac{\partial p}{\partial r},
  \end{equation}
is calculated to be 
\begin{equation}
  q_{\rm adv}^{\ -}(r,\ \xi) = \frac{\Re(r_0)}{2\sqrt{3}}\ \frac{B_0^2}{8\pi}
  \ \Omega_{\rm K}(r_0) \left( \frac{r}{r_0} \right)^{-4} \mbox{sech}^4 \xi 
\end{equation}
since the specific enthalpy is generally (i.e., irrespective of 
the polytropic law) given by $h=(5/2)(p/\rho)$ for an ideal gas. 
The result is exactly the same as the (effective) Joule heating rate, thus 
confirming a fully advective accretion flow. 

In addition to the heat generation discussed above, the pressure gradient 
along the flow also increases the enthalpy of the fluid. For a unit volume, 
this amounts to 
  \begin{equation}
      w(r,\ \xi) \equiv v_r\frac{\partial p}{\partial r} 
      = \frac{2\ \Re(r_0)}{\sqrt{3}}\ \frac{B_0^2}{8\pi}\ \Omega_{\rm K} (r_0) 
        \left( \frac{r}{r_0} \right)^{-4} \mbox{sech}^4 \xi
      = 4q_{\rm J}^{\ +}(r,\ \xi).
  \end{equation}
However, it is more convenient to discuss the energy budget in terms of the 
quantities per unit mass of a fixed fluid element. Then, the Bernoulli sum 
becomes of our interest. Neglecting higher order terms in $\Re^{-1}$ in 
the definition of the sum, we can show that 
\begin{equation}
  K \equiv \frac{v_{\varphi}^{\ 2}}{2} - \frac{GM}{r} + h = 0  
\end{equation}
from our solution. Generally the left-hand side may be a function of position, 
but in the present case it is constant throughout the flow. Therefore, in 
particular, $K$ is constant along the stream lines. This fact also implies 
that the flow is fully advective. The magnetic field plays only a catalytic 
role in the energy budget.

\section{INNER AND OUTER EDGES}

The ADAF solution described in the previous section naturally has a finite 
range of validity in the radial direction. The limiting radii for this 
validity are called the edges. The inner edge exists at a radius where 
the infall velocity becomes comparable with the rotation velocity. This 
means that, at around the inner edge, $\Re$ becomes of order unity and 
the assumption 5) in \S 2 becomes invalid. Therefore, we fix the inner 
edge at the radius where $\Re(r_{\rm in})=1$. As seen from equation 
(\ref{eqn:Re}), this guarantees the relation $\sigma^{-1}\propto\Delta^2$ 
and that $\Re(r)>1$ for $r>r_{\rm in}$. Thus the definition of the inner 
edge is conceptually clear, but the calculation of its explicit expression 
is generally not so simple. 

On the other hand, the definition of the outer edge is rather vague even 
in conceptual level. There is no general way of defining it. Both the 
definition of the outer edge and the explicit calculation of the inner 
edge depend on the configurations of external magnetic fields. Therefore, 
we discuss only two typical cases separately. Although similar problems have 
been discussed in K 87, we need to repeat and refine the discussions since 
the $r$-dependences of the present solution is slightly different from 
the previous one. 

The first example to be considered is a dipolar external field, because 
such a configuration may well be conceivable around a neutron star or a 
white dwarf in a mass exchanging binary system. In this case, all the 
advected energy is released finally at the surface of the central object. 
Anyway, from the $r^{-3}$-dependence of the external dipolar field, its 
strength exceeds that of the induced field in the region within the inner 
edge. Therefore the place at which the boundary value for $\tilde{b}_{\theta}$ 
should be fixed is the inner edge, i.e. $r_0=r_{\rm in}$, and the boundary 
value is 
\begin{equation}
  B_0 = \frac{\mu}{r_{\rm in}^{\ 3}},
\end{equation}
where $\mu$ is the magnitude of a dipole moment and its direction has been 
assumed to be vertical to the midplane of the accretion disk. Substituting 
this into equation (\ref{eqn:Mdot}), we obtain the explicit expression 
for $r_{\rm in}$ as 
\begin{equation}
  r_{\rm in} = \left( \frac{\Delta^2}{3}\ \frac{\mu^4}{GM\dot{M}^2} 
               \right)^{1/7}.
\end{equation}

The position of the outer edge is determined from a requirement of global 
magnetic-flux conservation. Namely, we require that the unperturbed 
magnetic flux penetrating the equatorial plane on the outside of the inner 
edge is compressed by the presence of an accretion flow within the region 
between the inner and outer edges, i.e.,
\begin{equation}
 B_0\int_{r_{\rm in}}^{\infty} \frac{r_{\rm in}^{\ 3}}{r^3}\ 2\pi r\ {\rm d}r 
     = \int_{r_{\rm in}}^{r_{\rm out}}\tilde{b}_{\theta}\ 2\pi r\ {\rm d}r. 
\end{equation}
In this case, therefore, the outer edge corresponds to a screening radius. 
The ratio of $r_{\rm out}$ to $r_{\rm in}$ is obtained from this equation 
as 
\begin{equation}
  \zeta \equiv \frac{r_{\rm out}}{r_{\rm in}} 
  = (1+\Delta^{-1})^2 \simeq \Delta^{-2}.
\label{eqn:dratio}
\end{equation}

The second example, which is of our present interest, is a uniform external 
field perpendicular to the equatorial plane. This is an idealization of 
the configurations expected in various AGNs. In this case, the external 
field dominates over the induced field at large radii. Therefore, the place 
at which the boundary value for $\tilde{b}_{\theta}$ should be fixed is the 
outer edge, i.e. $r_0=r_{\rm out}$. From equation (\ref{eqn:Mdot}) and 
the relation $\Re(r_{\rm out})=\zeta^{1/2}$, we have an implicit expression 
for $r_{\rm out}$ 
\begin{equation}
  r_{\rm out} = \left( \frac{3}{\Delta^2\zeta}\ \frac{GM\dot{M}^2}{B_0^{\ 4}} 
                \right)^{1/5}.
\label{eqn:Rout}
\end{equation}

Also in this case, the distance ratio of the outer to inner edges is 
obtained from a consideration of global magnetic-flux conservation. We 
assume that all the poloidal flux generated by the disk and penetrating 
the disk region should close in the region outside the outer edge. Then, 
the decrease in the flux of the external field in the outer region 
($r>r_{\rm out}$) is balanced by the increase in the inner region 
($r<r_{\rm in}$), i.e., 
\begin{equation}
  \pi r_{\rm in}^{\ 2}\ \tilde{b}_r(r_{\rm in}) 
    = \int_{r_{\rm in}}^{r_{\rm out}}\tilde{b}_{\theta}\ 2\pi r\ {\rm d}r,
\end{equation}
where the typical field strength in the inner region has been approximated 
by $\tilde{b}_r$. From this equation, we obtain the same result as 
(\ref{eqn:dratio}) again. Therefore, when $\Delta^2\ll 1$, equation 
(\ref{eqn:Rout}) reduces to 
\begin{equation}
  r_{\rm out} = \left( \frac{3GM\dot{M}^2}{B_0^{\ 4}} \right)^{1/5}.
\label{eqn:aRout}
\end{equation}

Here, we briefly return to the energy budget in the disk and discuss it, 
this time, from a global point of view. It has been shown in the previous 
section that the rate of increase of the enthalpy in a unit volume through 
the current dissipation and the plasma compression is 
$5q_{\rm J}^{\ +}(r,\ \xi)$. Integrating $q_{\rm J}^{\ +}$ over the whole 
volume of the disk region, we obtain 
\begin{equation}
  Q_{\rm J} = \int_{r_{\rm in}}^{r_{\rm out}}{\rm d}r 
    \int_{-\infty}^{\infty}{\rm d}\xi\ 2\pi\Delta r^2 q_{\rm J}^{\ +}(r,\ \xi) 
  = \frac{GM\dot{M}}{6} \left[ \frac{1}{r_{\rm in}} 
    - \frac{1}{r_{\rm out}} \right],
\end{equation}
where $B_0$ has been eliminated by equation (\ref{eqn:Mdot}). The result 
depends neither on the type of the external field discussed above nor on 
the explicit expressions for $r_{\rm in}$ and $r_{\rm out}$. As expected 
for a fully advective flow, the increase of the enthalpy in the whole disk 
region is exactly balanced by the net enthalpy flux coming out of the 
disk region: 
\begin{equation}
  F_{\rm enth} \equiv \int h\rho{\bf v}\cdot {\rm d}{\bf S} 
  = \frac{5GM\dot{M}}{6} \left[ \frac{1}{r_{\rm in}} 
    - \frac{1}{r_{\rm out}} \right] 
  = 5Q_{\rm J},
\end{equation}
where d{\bf S} is the surface element of the disk region. 

On the other hand, the kinetic energy flux coming out of the disk region is 
\begin{equation}
  F_{\rm rot} = \frac{\dot{M}}{2} \left[ v_{\varphi}^{\ 2}(r_{\rm in}) 
          - v_{\varphi}^{\ 2}(r_{\rm out}) \right] 
      = \frac{GM\dot{M}}{6} \left[ \frac{1}{r_{\rm in}} 
        - \frac{1}{r_{\rm out}} \right].
\end{equation}
Thus, it has turned out that the net input of the gravitational energy 
in the disk region per unit time, 
$-GM\dot{M}\ (r_{\rm in}^{\ -1} - r_{\rm out}^{\ -1})$, is converted into 
the kinetic energy of quasi-Keplerian rotation (by 1/6) and the enthalpy 
which has been added as heat (by 1/6) and also as compression (by 2/3).

\section{SCALING FOR ACTIVE GALACTIC NUCLEI}

Since we are now interested in the AGN activities, the quantities in our 
model are suitably scaled here to such circumstances. The model is 
completely specified by three boundary values and one parameter, i.e., 
mass of the central black hole $M$, mass accretion rate $\dot{M}$, 
strength of the global magnetic field $B_0$ and the dissipation parameter 
$\Delta$. They are normalized, respectively, by $10^8 M_{\odot}$, the 
Eddington accretion rate $\dot{M}_{\rm E}$, 1 Gauss and 0.1: 
  \begin{equation}
    m \equiv \frac{M}{10^8M_{\odot}}, \quad \dot{m} 
    \equiv \frac{\dot{M}}{\dot{M}_E}, \quad b_0 \equiv \frac{B_0}{1{\rm G}}, 
    \quad \delta \equiv \frac{\Delta}{0.1}.
  \end{equation} 
The explicit expression 
for the Eddington accretion rate is 
\begin{equation}
  \dot{M}_{\rm E} = 1.4\times10^{25} 
      \left( \frac{M}{10^8M_{\odot}} \right) \quad \mbox{g\ s$^{-1}$}.
\end{equation}
Since the reference radius should be fixed at the outer edge, i.e. 
$r_0=r_{\rm out}$, we have $\Re(r_{\rm out})=\Delta^{-1}$. 

In terms of the above non-dimensional quantities, the boundary values of 
other quantities appearing in the model are expressed as 
\begin{equation}
   r_{\rm out} = 0.96\times10^{17}\ b_0^{-4/5}\ \dot{m}^{2/5}\ m^{3/5} 
      \quad \mbox{cm},
\end{equation}
\begin{equation} 
  \tilde{b}_{\varphi}(r_{\rm out}) = 10\ \delta^{-1}\ b_0 \quad \mbox{G},
\end{equation}
\begin{equation}
  v_{\rm K}(r_{\rm out}) = 3.7\times10^8\ b_0^{2/5}\ \dot{m}^{-1/5}\ m^{1/5} 
     \quad \mbox{cm\ s$^{-1}$},
\end{equation}
\begin{equation}
  \tilde{p}(r_{\rm out}) = 4.0\ \delta^{-2}\ b_0^2 \quad 
    \mbox{dyne\ cm$^{-2}$},
\end{equation}
\begin{equation}
  T(r_{\rm out}) = 2.7\times10^8\ b_0^{4/5}\ \dot{m}^{-2/5}\ m^{2/5} 
      \quad \mbox{K},
\end{equation}
\begin{equation}
  \tilde{\rho}(r_{\rm out}) = 8.8\times10^{-17}\ \delta^{-2}\ b_0^{6/5}
      \ \dot{m}^{2/5}\ m^{-2/5} \quad \mbox{g\ cm$^{-3}$}.
\end{equation}

All other induced quantities can be calculated from the set of analytic 
solutions given in \S 4 and their boundary values given above. 
For example, integrating the density over the disk height, we obtain 
the surface density which becomes independent of $r$ in our model: 
\begin{displaymath}
 \Sigma(r) \equiv \int_{-\infty}^{\infty}\rho(r,\ \xi)r\Delta
      \ {\rm d}\xi = \mbox{const.} \equiv \Sigma(r_{\rm out}),
\end{displaymath}
\begin{equation}
 \Sigma(r_{\rm out}) = 1.7\times10\ \delta^{-1}\ b_0^{2/5}\ \dot{m}^{4/5}
      \ m^{1/5} \quad \mbox{g\ cm$^{-2}$}.
\end{equation}
The dependence of $\Sigma \propto \dot{M}^{4/5}$ suggests that the disk is 
secularly stable (e.g., Kato, Fukue \& Mineshige 1998). Further integrating 
$\Sigma(r)$ over the radius, we have the total mass in the disk as 
\begin{equation}
  M_{\rm D} \simeq \pi r_{\rm out}^{\ 2}\Sigma 
    = 4.9\times 10^{35}\ \delta^{-1}\ b_0^{-6/5}\ \dot{m}^{8/5}\ m^{7/5} 
    \quad \mbox{g}.
\end{equation}

In order for our analytic solution to serve well as a model of ADAF in 
a global magnetic field, the radiation energy which is expected to come out 
of the disk should be really negligible in the physical situations of our 
interest. This point will be checked here. The opacity due to the electron 
scattering is $\kappa_{\rm es} = 0.4$ cm$^2$\ g$^{-1}$ and the free-free 
opacity is written as $\kappa_{\rm ff} = 0.6\times10^{23}\ \rho\ T^{-7/2}$ 
cm$^2$\ g$^{-1}$. 
In view of the scaled values for $\rho$ and $T$, one can safely conclude 
that the former opacity is dominant over the latter. Then, the optical depth 
of the disk becomes $r$-independent: 
\begin{displaymath}
  \tau(r) = \int_{-\infty}^{\infty}\kappa(r,\ \xi)\ \rho(r,\ \xi)\ r\Delta
  \ {\rm d}\xi \simeq \kappa_{\rm es}\Sigma(r_2)
\end{displaymath}
\begin{equation}
  = 3.4\ \delta^{-1}\ b_0^{2/5}\ \dot{m}^{4/5}\ m^{1/5}.
\end{equation}
Provided that $\dot{m}\ll 1$, it can be confirmed from this expression 
that the disk is everywhere optically thin unless its opening angle is 
extremely small. 

As a representative of radiation losses, we evaluate the Bremsstrahlung 
from the disk. This is because it can be evaluated by a simple formula 
and other mechanism such as the synchrotron loss seems to have a similar 
order of magnitude (Narayan et al. 1995, Manmoto et al. 1997, Narayan 
et al. 1998). The radiation flux from the disk surface is 
\begin{displaymath}
  F_{\rm B}^{\ -}(r) 
  = \int_{-\infty}^{\infty}q_{\rm B}(r,\ \xi)\ r\Delta\ {\rm d}\xi  
  = F_{\rm B}^{\ -}(r_{\rm out}) \left( \frac{r}{r_{\rm out}} \right)^{-3/2},
\end{displaymath}
\begin{equation}
  F_{\rm B}^{\ -}(r_{\rm out}) = 2.0\times10^7\ \delta^{-3}\ b_0^2
  \ \dot{m}\ m \quad \mbox{ergs\ cm$^{-2}$\ s$^{-1}$},
\label{eqn:Bloss}
\end{equation}
where $q_{\rm B}(r,\ \xi)$ is the volume emissivity. On the other hand, 
height integrating the rate of enthalpy increase, we obtain for energy 
input rate at a radius $r$ as 
\begin{displaymath}
  Q_{\rm enth}^{\ +}(r) = \int_{-\infty}^{\infty} 
    [q_{\rm J}(r,\ \xi) + w(r,\ \xi)]\ r\Delta\ {\rm d}\xi  
  = Q_{\rm enth}^{\ +}(r_{\rm out}) \left( \frac{r}{r_{\rm out}} 
    \right)^{-3/2},
\end{displaymath}
\begin{equation}
  Q_{\rm enth}^{\ +}(r_{\rm out}) 
  = 2.9\times10^7\ b_0^{12/5}\ \dot{m}^{-1/5}\ m^{1/5} 
  \quad \mbox{ergs\ cm$^{-2}$\ s$^{-1}$}.
\end{equation}
Comparing this with the expression (\ref{eqn:Bloss}), we can see that 
the radiation loss is actually negligible under the circumstances with 
$\dot{m}\ll 1$ unless the disk is extremely thin. 

Although our model assumes, for simplicity, a common temperature for 
both the electron and ion components, actually it may take different values 
for different components as is the case in the viscous ADAF models. 
Such a difference in temperature, however, may be less important in the 
resistive ADAFs because the (effective) Joule dissipation will heat 
mainly the electron component (Bisnovati-Kogan \& Lovelace 1997) which 
is easier to cool down by radiation than the ion component. In any case, 
that is beyond the scope of the present paper.

\section{DISCUSSION}

In this section, we discuss main issues which may be raised about 
our model, introducing speculations to some extent. 
The first is about the accuracy of our set of analytic solutions. 
As described in \S 4, the separation of variables are performed only 
approximately. 
It is accurate at the midplane of a disk, but is invalid for large 
$\xi$. The errors in the angular dependence may exceed 50\% for 
$\vert\xi\vert>0.5$, and certain inconsistencies appear beyond 
$\tanh\vert\xi\vert=1/\sqrt{3}$ where (d$^3$/d$\xi^3$)\ tanh$\xi$ 
changes its sign. 
One may therefore identify there (i.e., $\vert\xi\vert\sim 1$) as 
the surfaces of a disk. 
Within this range, the set of solutions serves well for most of 
our purposes because we are usually interested not in the exact 
angular dependences but in their qualitative behavior. 

Angular momentum is expected to be carried away from the disk 
by magnetic stresses along the externally given poloidal 
magnetic lines of force. 
In the case of a dipole type external field it is transferred to 
the central star, and in the case of a uniform external field, 
to distant regions. 
The rate of angular momentum extraction from the upper and 
lower surfaces of a disk can be calculated from our solution. 
The result is proportional to 
$\dot{M}[\sqrt{GMr_{\rm out}} - \sqrt{GMr_{\rm in}}]$, 
and the proportionality constant depends on the definition of 
the surfaces $\xi=\pm\xi_0$. Such an ambiguity is also a 
reflection of the approximate nature of the solution. 

Although a definite answer cannot be obtained in the present 
stage of approximation, the above picture for angular momentum 
extraction seems to be physically plausible also from a global 
consideration of the accretion process. 
A schematic picture is shown in Fig.\ 1 for a uniform external 
field. 
This field is an idealization of such a large-scale field of 
galactic origin whose sources are in distant regions. 
The rotation of the accreting plasma in this poloidal field 
acts as a dynamo in a DC-circuit and drives a radial current 
in the disk. 
We expect this current to close its circuit around distant 
regions and reaching to the polar regions (as for the supply 
of plasma to this regions, see below). 
The presence of such a globally circulating poloidal current 
system guarantees the generation of toroidal magnetic field outside 
the disk, which is necessary for the transfer of angular momentum 
by magnetic stresses. 

Thus, the angular momentum can be extracted through a nearly 
vacuum space, but we do not intend to deny the possibility 
of the extraction also by wind type outflows (e.g., Blandford 
\& Payne 1982). 
Rather, an accretion flow with a divergent component in the 
$\theta$-direction has been obtained also in our scheme when the 
simplifying assumption of $v_{\theta}=0$ is relaxed (Kaburaki 1987). 

The ${\bf j}_{\rm p}\times {\bf b}_{\varphi}$ force acting on 
the poloidal current ${\bf j}_{\rm p}$ whose origin has been 
discussed above contributes to the confinement of accreting plasma 
towards the equatorial plane and, possibly, also to the collimation 
and acceleration of a bipolar jet system (see, Kaburaki \& Itoh 1987). 
The former effect has been described in the present paper as the 
magnetic confinement of an accretion disk (the $\theta$-component 
of the equation of motion). 
The latter two effects are interpreted as the pinch effect on the 
return current near the polar axis and the acceleration by magnetic 
pressure-gradient force due to $b_{\varphi}$, respectively. 

Therefore, the development of a toroidal magnetic field in the 
middle-latitude regions seems to naturally explain such an 
association of disk-like structure and bipolar-jet structure 
as is frequently observed in many astrophysical situations. 
Interestingly enough, essentially the same story as the above 
for the formation of a disk-like structure and for the collimation 
and acceleration of a jet applies even to the case of disk-like 
outflows which are expected to exist around rapidly rotating, 
magnetized stars including pulsars (Kaburaki 1989).  
This is because the sense of 
the ${\bf j}_{\rm p}\times {\bf b}_{\varphi}$ force does not change 
even in that case, in spite of the change in the sense of $b_{\varphi}$ 
(and therefore that of ${\bf j}_{\rm p}$) which is required  to 
realize an outflow (i.e., angular momentum should be added to the disk). 
Thus, the disk-jet association is understood naturally in the 
same scheme even in the case of outflows. 

The source of plasmas in the polar regions is of course an accretion 
disk. 
Within the inner edge of an accretion disk, there develops 
a strong poloidal magnetic field swept by the accretion flow. 
This field may form a core of a bipolar-jet system. 
The accretion flow would be decelerated within the inner edge by 
the strong magnetic field perpendicular to it and, although main fraction 
of it may be eventually swallowed by the central black hole, some part 
may be turned its direction to follow the core field and then accelerated 
along it (see Fig.\ 1).

It may seem at first rather curious that, as far as our solution 
is concerned, no energy is extracted from the disk associated with 
the angular momentum extraction by magnetic stresses. 
As confirmed by substituting our solution into the leading-order 
expressions in $\Delta$ of the electric field, 
\begin{equation}
  E_r = \frac{c}{4\pi\sigma\Delta}\frac{1}{r}
        \frac{\partial b_{\varphi}}{\partial\xi} 
        + \frac{1}{c}v_{\varphi}b_{\theta}, \quad 
  E_{\theta} = -\frac{1}{c}(v_{\varphi}b_r - v_r b_{\varphi}), \quad
  E_{\varphi} = 0,
\end{equation}
we have ${\bf E}=0$ in the disk and therefore the vanishing Poynting 
flux. 
The situation, however, is quite analogous to that in the viscous 
ADAF models in which angular momentum is transported radially outwards 
by viscous stresses but energy is dissipated locally in the flow. 

The vanishing of the Poynting flux is a consequence of idealization 
that the external electric loads in which some part of energy may be 
consumed is negligible in determining the structure of an accretion 
flow. 
In such a situation, there is no need to transport energy to the outside. 
This is an opposite extreme of ideal MHD disks. 
Since no energy dissipation can occur in an ideal MHD disk, the whole 
power input is carried away in terms of the Poynting flux to distant 
regions where the dissipation is assumed to occur. 
In view of the above discussion of global configurations, at least, 
the jet should be included as an external load in a more realistic 
determination of the current in the disk. 
However, of course this is beyond the scope of the present work. 

The global MHD stability of accretion disks in the resistive ADAF 
model may be an important issue. 
This is because the ratio of the toroidal to the poloidal magnetic 
fields which is represented by $\Re$ is assumed to be large 
($\Re>1$ and $\Re^2\gg 1$) in the model, while the configurations 
with too large $\Re$ are generally believed to be unstable. 
Since $\Re$ increases from $\sim 1$ at the inner edge like $r^{1/2}$ 
towards the outer edge ($\Re(r_{\rm out})=\Delta^{-1}$ in the case 
of uniform external field), the outermost regions are apt to be 
unstable to the instabilities of helical type. 
Although the determination of the critical value for $\Re$ in actual 
resistive ADAFs is again beyond the scope of this paper, it can be 
said qualitatively that only not so widely extended disks and not so 
thin disks are expected to be maintain safely. 

As for the local MHD stabilities, we have already performed an 
analysis based on the background solution obtained in the present 
paper and have found that there are actually growing modes. 
Although further investigations including non-linear evolutions 
are needed for a definite conclusion, the instabilities are expected 
to be not so strong as to destroy the whole structure. 
Rather, they may cause a turbulence in the flow which is the origin 
of some types of anomalous resistivity or viscosity (e.g., Balbus 
\& Hawley 1991). 
The results of the local analysis will be reported elsewhere.

\appendix
\section{APPENDIX}

Although a reduced Keplerian velocity has appeared in our resistive 
ADAF model, the rotation law may be different for different geometries of 
the accretion flows. Here, this point will be demonstrated by considering 
two typical cases: one is the disk-like flow of constant height (i.e., 
$H=\mbox{const.}$) and the other is that of constant opening angle (i.e., 
$\Delta=\mbox{const.}$). For simplicity, we assume that the disk structure 
is maintained only by the gravity and that the infall velocity is negligibly 
small compared with the rotation velocity. 

When the disk height is constant, the pressure force is almost vertical 
to the equatorial plane and it is convenient to adopt cylindrical coordinates 
($\varpi$, $\varphi$, $z$). The equations of force balance in a poloidal 
plane are, therefore, 
\begin{equation}
  \frac{GM}{r^2}\ \frac{\varpi}{r} = \frac{v_{\varphi}^{\ 2}}{\varpi},
 \label{eqn:A1}
\end{equation}
\begin{equation}
  \frac{GM}{r^2}\ \frac{z}{r} 
   = \frac{1}{\rho}\ \frac{\partial p}{\partial z},
 \label{eqn:A2}
\end{equation}
where $r=(\varpi^2+z^2)^{1/2}$ and the pressure term in the 
$\varpi$-component has been neglected. Equation (\ref{eqn:A2}) implies a 
gravitational confinement of the accreting plasma. 

Equation (\ref{eqn:A1}) yields the Kepler {\it angular velocity} 
\begin{equation}
  \Omega(\varpi,\ z) = \left( \frac{GM}{r^3} \right)^{1/2} 
    \equiv \Omega_{\rm K}(r). 
 \label{eqn:A3}
\end{equation}
However, the plane of rotation is always parallel to the equatorial plane, 
so that the center of rotation does not coincide with the center of 
gravitational attraction except for the rotation in the disk's midplane. 
Since the angular velocity is Keplerian, the centrifugal force is generally 
smaller than the gravity at an arbitrary $r$: 
\begin{equation}
  \frac{v_{\varphi}^2}{\varpi} = \frac{GM}{r}\sin\theta \leq \frac{GM}{r}. 
\label{eqn:A4}
\end{equation}
From equation (\ref{eqn:A2}) we have a rough estimate for the ratio of 
sound velocity to the Kepler velocity: 
\begin{equation}
  \frac{C_{\rm S}^{\ 2}}{v_{\rm K}^{\ 2}} \sim \frac{H^2}{r^2}.
\label{eqn:A5}
\end{equation}
Since $H$ is a constant, this suggests a temperature variation of the 
form $T(r)\propto r^{-3}$. 

When the opening angle of a disk is constant, the pressure force is 
almost along the $\theta$-direction and it is convenient to adopt 
spherical polar coordinates ($r$, $\theta$, $\varphi$). The force balance 
equations are 
\begin{equation}
  \frac{v_{\varphi}^{\ 2}}{r} = \frac{GM}{r^2},
\label{eqn:A6}
\end{equation}
\begin{equation}
  \frac{v_{\varphi}^{\ 2}}{r}\cot\theta 
     = \frac{1}{\rho r}\ \frac{\partial p}{\partial\theta},
\label{eqn:A7}
\end{equation}
where the pressure term in $r$-component has been neglected compared 
with the gravity. Equation (\ref{eqn:A7}) implies a centrifugal confinement 
of the accreting plasma. 

Equation (\ref{eqn:A6}) yields the Kepler {\it rotational velocity}
\begin{equation}
  v_{\varphi}(r,\ \theta) = \left( \frac{GM}{r} \right)^{1/2} 
     \equiv v_{\rm K}(r). 
\label{eqn:A8}
\end{equation}
The plane of rotation is again parallel to the equatorial plane, so that 
the center of rotation does not coincides with the center of gravitational 
attraction except for the rotation in the disk's midplane. Since the 
rotational velocity is Keplerian, the centrifugal force generally exceeds 
the gravity at an arbitrary $r$: 
\begin{equation}
  \frac{v_{\varphi}^2}{\varpi} = \frac{GM}{r}\ \frac{1}{\sin\theta}
    \geq \frac{GM}{r}. 
\label{eqn:A9}
\end{equation}
From equation (\ref{eqn:A7}) we have a rough estimate for the ratio of 
sound velocity to the Kepler velocity: 
\begin{equation}
  \frac{C_{\rm S}^{\ 2}}{v_{\rm K}^{\ 2}} 
     \sim \frac{\vert\cot\Delta\vert}{\Delta}.
\label{eqn:A10}
\end{equation}
Since $\Delta$ is a constant, this suggests a temperature variation of 
the form $T(r)\propto r^{-1}$. 

From the above discussions, we can see that the geometry of disks 
has essential effects on their properties. 
The rotation law is different between the disks of constant $H$ and 
constant $\Delta$. 
The radial dependences of the temperature is also quite different. 
The same dependence as the virial temperature is obtained only in 
the disks of constant $\Delta$, but thin disks can be realized 
there only when the force of confinement other than gravity is present 
(i.e., equation (\ref{eqn:A7}) cannot hold when $\Delta \ll 1$ 
even if the temperature is nearly virial). 
In the case of constant $H$, on the other hand, thin disks are 
realized if the temperature is much smaller than the virial one.


\newpage  
\figcaption  
{Schematic drawing of the global geometries of magnetic 
field and plasma flow. A poloidally circulating current system 
(${\bf j}_p$) driven by the rotational motion of accreting plasma 
generates a toroidal magnetic field $b_{\varphi}$ in addition to 
a nearly uniform external field. 
The presence of this toroidal field outside the disk guarantees 
the magnetic extraction of angular momentum from the disk. 
This field acts to confine the accreting plasma towards the 
equatorial plane and also has a tendency to collimate and accelerate 
the plasma in the polar regions. 
If the condition is favorable, the plasma in the polar regions 
may form a set of bipolar jets.}


\newpage

\begin{references}

\reference{1}
{Abramowicz, M., Chen, X., Kato, S., Lasota, J.\ P., \& Regev, O. 
1995, ApJ, 438, L37}
\reference{1}
{Balbus, S.\ A., \& Hawley J.\ F. 1991 ApJ 376, 214}
\reference{1}
{Bisnovati-Kogan, G.\ S., \& Lovelace, R.\ V.\ E. 1997, ApJ, 486, L43}
\reference{1}
{Blandford, R.\ D., \& Payne, D.\ G. 1982, MNRAS, 199, 883}
\reference{1}
{Ichimaru, S. 1977, ApJ, 214, 840}
\reference{1} 
{Kaburaki, O. 1986, MNRAS, 220, 321}
\reference{1}
{Kaburaki, O. 1987, MNRAS, 229, 165} 
\reference{1}
{Kaburaki, O. 1989, MNRAS, 237, 49}
\reference{1}
{Kaburaki, O. 1999, in Disk Instabilities in Close Binary Systems, 
ed. S.\ Mineshige \& J.\ C.\ Wheeler (Tokyo: Universal Academy Press), 325} 
\reference{1}
{Kaburaki O., \& Itoh M. 1987, A\&Ap, 172, 191}
\reference{1}
{Kato, S., Fukue, J., \& Mineshige, S. 1998, Black-Hole Accretion Disks 
(Kyoto: Kyoto Univ. Press)}
\reference{1}
{Manmoto, E., Mineshige, S., \& Kusunose, M. 1997, ApJ, 489, 791}
\reference{1}
{Narayan, R., \& Yi, I. 1994, ApJ, 428, L13}
\reference{1}
{Narayan, R., \& Yi, I. 1995, ApJ, 453, 710}
\reference{1}
{Narayan, R., Yi, I., \& Mahadevan R. 1995, Nature, 374, 623}
\reference{1}
{Narayan, R., Mahadevan, R., Gridlay, J.\ E., Popham, P.\ G., \& Gammie, C. 
1998, ApJ, 492, 554}
\reference{1}
{Narayan, R., Mahadevan, R., \& Quataert, E. 1999, in The Theory of Black 
Hole Accretion Discs, ed. M.\ A.\ Abramowicz, G.\ Bjornsson, \& 
J.\ E.\ Pringle (Cambridge: Cambridge U.\ P.) in press}
\reference{1}
{Rees, M.\ J., Begelman, M.\ C., Blandford, R.\ D., \& Phinney, E.\ S. 
1982, Nature, 295, 17}
\reference{1}
{Shakura, N.\ I., \& Sunyaev R.\ A. 1973, A\&Ap, 24, 337}
\reference{1}
{Yusef-Zadeh, F., Morris, M., \& Chance, D. 1984, Nature, 310, 557}

\end{references}
\end{document}